\documentclass[aps,prb,superscriptaddress,showpacs,twocolumn,bibnotes]{revtex4}
\usepackage{graphicx}
\usepackage{amssymb}
\usepackage{amsmath}

\newcommand{\bk}{{\bf k}}
\newcommand{\bq}{{\bf q}}

\newcommand{\bK}{{\bf K}}

\newcommand{\bdelta}{{\boldsymbol\delta}}
\newcommand{\bsigma}{{\boldsymbol\sigma}}

\renewcommand{\Im}{{\mathop{\rm{Im}}\nolimits\,}}
\renewcommand{\Re}{{\mathop{\rm{Re}}\nolimits\,}}

\newcommand{\diag}{{\mathop{\rm{diag}}\nolimits\,}}

\newcommand{\kB}{k_{\mathrm{B}}}
\newcommand{\kF}{k_{\mathrm{F}}}

\newcommand{\vF}{v_{\mathrm{F}}}

\begin{document}

\title{Linear response correlation functions in strained graphene}

\author{F. M. D. Pellegrino}
\affiliation{Dipartimento di Fisica e Astronomia, Universit\`a di Catania,\\
Via S. Sofia, 64, I-95123 Catania, Italy}
\affiliation{CNISM, UdR Catania, I-95123 Catania, Italy}
\author{G. G. N. Angilella}
\affiliation{Dipartimento di Fisica e Astronomia, Universit\`a di Catania,\\
Via S. Sofia, 64, I-95123 Catania, Italy}
\affiliation{CNISM, UdR Catania, I-95123 Catania, Italy}
\affiliation{Scuola Superiore di Catania, Universit\`a di Catania,\\
Via Valdisavoia, 9, I-95123 Catania, Italy}
\affiliation{INFN, Sez. Catania, I-95123 Catania, Italy}
\author{R. Pucci}
\affiliation{Dipartimento di Fisica e Astronomia, Universit\`a di Catania,\\
Via S. Sofia, 64, I-95123 Catania, Italy}
\affiliation{CNISM, UdR Catania, I-95123 Catania, Italy}

\date{\today}

\begin{abstract}
After deriving a general correspondence between linear response correlation
functions in graphene with and without applied uniaxial strain, we study the
dependence on the strain modulus and direction of selected electronic
properties, such as the plasmon dispersion relation, the optical conductivity,
as well as the magnetic and electric susceptibilities. Specifically, we find
that the dispersion of the recently predicted transverse plasmon mode exhibits
an anisotropic deviation from linearity, thus facilitating its experimental
detection in strained graphene samples.
\medskip
\pacs{%
73.20.Mf, 
62.20.-x, 
81.05.ue 	
}
\end{abstract} 

\maketitle

\section{Introduction}

Graphene is a truly two-dimensional (2D) electronic system, based on an
atomically thin carbon honeycomb lattice \cite{Novoselov:05a}. Low-energy
quasiparticles can be described as massless Dirac fermions, with a cone
dispersion relation in reciprocal space around the so-called Dirac points $\bK$,
$\bK^\prime$, and a linearly vanishing density of states (DOS) at the Fermi
level \cite{CastroNeto:08,Abergel:10}. Such a linear spectrum and reduced
dimensionality yield remarkable behaviors already in the non-interacting limit
of several electronic properties of graphene. These include, \emph{inter alia,}
the reflectivity \cite{Nair:08}, the optical conductivity
\cite{Kuzmenko:08,Wang:08,Mak:08,Stauber:08a,Pellegrino:09b}, the plasmon dispersion relation
\cite{Hwang:07a,Polini:09,Pellegrino:10a,Pellegrino:10c}, as well as a newly
predicted transverse electromagnetic mode \cite{Mikhailov:07}, which is
characteristic of a 2D system with a double band structure, such as graphene.
Moreover, the relevance of lossless plasmons in graphene in the infrared
frequency range has been emphasized, with possible applications in nanophotonics
\cite{Li:08,Jablan:09}. These properties can be extracted from the study of the
appropriate correlation functions within linear response theory
\cite{Wunsch:06,Stauber:10a,note:Basko}.

Within the Dirac approximation, rotational invariance implies thet the
current-current correlation function may be decomposed in a longitudinal and a
transverse contribution, the former being related to the density-density
correlation function via the continuity equation \cite{Stauber:10a,Principi:09}.
Moreover, in the case of massless Dirac fermions, it has been observed that the
current-current response is simply proportional to its pseudospin-pseudospin
counterpart \cite{Principi:09}. On the other hand, the magnetic susceptibility
is related to the transverse contribution \cite{Stauber:10a,Ando:06,Koshino:09}.
Specifically, the noninteracting Dirac model yields an orbital magnetic
susceptibility $\chi_m (q\to0) \propto \delta(\mu)$, in the long wavelength
limit \cite{Principi:09}. This is consistent with earlier results for graphite
\cite{McLure:56}, obtained using Wallace's two-dimensional band structure for a
graphene layer \cite{Wallace:47}. Such a finding would predict no response to a
uniform, static magnetic field, away from half-filling. This is of course
partially compensated by a smearing of the $\delta$-function at finite
temperatures, already in the noninteracting limit. Still at zero temperature and
in the noninteracting limit, one recovers a nonzero magnetic response also away
from half-filling, when the honeycomb lattice structure is considered
\cite{GomezSantos:11}. The effect of the interactions has been considered in
Ref.~\onlinecite{Principi:10}, where it is shown that an interacting 2D Dirac
electron liquid develops a magnetic response also at finite doping.  Concerning
the response to an external electromagnetic field, graphene is also unique among
other conventional 2D electron systems, in that it has been predicted that it
can sustain a transverse plasmon mode \cite{Mikhailov:07}, as a consequence of
its double band structure. More recently, such a transverse mode has been
predicted also for bilayer graphene \cite{Jablan:11a}.

Here, we will consider the effect of strain on the various electronic properties
that may be described by linear response correlation functions. Indeed, a
deformation of the lattice through the application of uniaxial strain or
hydrostatic pressure is expected to produce modifications also in the electronic
structure of graphene. Recently, it has been proposed that nanodevices based on
graphene could be engineered on the basis of the expected strain-induced
modifications of the deformed graphene sheet (origami electronics)
\cite{Pereira:09}. This is made possible by the exceptional mechanical
properties of graphene, as is the case for other carbon compounds. For instance,
despite its reduced dimensionality, graphene is characterized by a sizeable
tensile strength and stiffness \cite{Booth:08}, with graphene sheets being
capable to sustain elastic deformations as large as $\approx 20$\%
\cite{Kim:09,Liu:07,Cadelano:09,Choi:10,Jiang:10}. Larger strains would then
induce a semimetal-to-semiconductor transition,
with the opening of an energy gap \cite{Gui:08,Pereira:08a,Ribeiro:09,Cocco:10},
and it has been demonstrated that such an effect critically depends on the
direction of applied strain \cite{Pellegrino:09b,Pellegrino:09c}.

The paper is organized as follows. In Sec.~\ref{sec:model} we present our
formalism for treating graphene under uniaxial strain, and derive our central
result for a generic linear response function for a deformed graphene sheet.
This, in particular, applies to the density-density and current-current
correlation functions. In Sec.~\ref{sec:polarization} we then study the electron
polarization, with emphasis on the strain dependence of the plasmon dispersion
relation and the conductivity. Specifically, we find a strain-induced
anisotropic enhancement of the deviations from linearity of the transverse
plasmon \cite{Mikhailov:07}, which should facilitate its experimental detection.
In Sec.~\ref{sec:susceptibilities} we derive the strain dependence of the
magnetic and electric susceptibilities. Finally, we summarize our results in
Sec.~\ref{sec:conclusions}.

\section{Model}
\label{sec:model}

The low-energy Hamiltonian for noninteracting quasiparticles around a Dirac
point, say $\bK$, has the well-known linear form \cite{CastroNeto:08}
\begin{equation}
H^{(0)} = \hbar \vF \bsigma \cdot \bq,
\label{eq:Hunstrained}
\end{equation}
where $\vF$ is the Fermi velocity, $\bsigma = (\sigma_x , \sigma_y)$, with
$\sigma_x$, $\sigma_y$ Pauli matrices, and $\bq$ is the wavevector displacement
from the Dirac point one is referring to, \emph{i.e.} $\bk = \bK + \bq$. Here
and below, a superscript zero denotes absence of strain. The effect of strain is
then that of modifying the lattice vectors as $\bdelta_\ell = (\mathbb{I} + 
{\boldsymbol\varepsilon}) \cdot \bdelta^{(0)}_\ell$ ($\ell=1,2,3$), where
$\bdelta_1^{(0)} = a(\sqrt{3},1)/2$,  $\bdelta_2^{(0)} = a(-\sqrt{3},1)/2$, 
$\bdelta_3^{(0)} = a(0,-1)$ are the relaxed (unstrained) vectors connecting two
nearest-neighbor (NN) carbon sites, with $a=1.42$~\AA, the equilibrium C--C
distance in a graphene sheet \cite{CastroNeto:08}, and
${\boldsymbol\varepsilon}$ is the strain tensor \cite{Pereira:08a}
\begin{equation}
{\boldsymbol\varepsilon} = 
\frac{1}{2} \varepsilon [(1-\nu){\mathbb I} + (1+\nu) A(\theta)],
\label{eq:strainmat}
\end{equation}
where
\begin{equation}
A (\theta) = \sigma_z e^{2i\theta\sigma_y} 
= \cos(2\theta) \sigma_z + \sin(2\theta) \sigma_x .
\end{equation}
In Eq.~(\ref{eq:strainmat}), $\theta$ denotes the angle along which the strain
is applied, with respect to the $x$ axis in the lattice coordinate system,
$\varepsilon$ is the strain modulus, and $\nu$ is Poisson's ratio. While in the
hydrostatic limit $\nu=-1$ and ${\boldsymbol\varepsilon} = \varepsilon{\mathbb
I}$, in the case of graphene one has $\nu=0.14$, as determined from \emph{ab
initio} calculations \cite{Farjam:09}, to be compared with the known
experimental value $\nu=0.165$ for graphite \cite{Blakslee:70}. The special
values $\theta=0$ and $\theta=\pi/6$ refer to strain along the zig~zag and
armchair directions, respectively.

The overall effect of a moderately low applied uniaxial strain on the low-energy
Hamiltonian is that of shifting the location of the Dirac point in momentum
space as $\bK \to \bk_D$, and changing the shape of the Dirac cone, into a
deformed one, with elliptical section \cite{Pellegrino:09b}. Such a picture
applies for strain moduli below the value at which a gap opens in the energy
spectrum, which takes place at $\varepsilon\simeq 20$\%
\cite{Kim:09,Liu:07,Cadelano:09,Choi:10,Jiang:10}. In particular, setting $\bk =
\bk_D + \bq$, with $\bq$ measuring now the vector displacement from the shifted
Dirac point, the Fermi velocity, defined as the slope of the Dirac cone in the
direction of $\bq$, will now have anisotropic components $c_\parallel \vF$,
$c_\perp \vF$ along the direction of applied strain and the direction orthogonal
to it, respectively, with
\begin{subequations}
\begin{eqnarray}
c_\parallel &=& 1-2\kappa\varepsilon ,\\
c_\perp &=& 1+2\kappa\nu\varepsilon,
\end{eqnarray}
\end{subequations}
where $\kappa = (a/2t)|\partial t/\partial a| -\frac{1}{2} \approx 1.1$ is
related to the logarithmic derivative of the nearest-neighbor hopping $t$ at
$\varepsilon=0$. Thus, the low-energy Hamiltonian around $\bk_D$ maintains a
linear form even in the presence of strain, and can still be written as
\begin{equation}
H = \hbar \vF  \bsigma \cdot \bq^\prime ,
\label{eq:Hstrained}
\end{equation}
where now
\begin{equation}
\bq^\prime = R(\theta) S(\varepsilon) R(-\theta) \bq ,
\label{eq:straintransform}
\end{equation}
with $R(\theta)$ the rotation matrix in the direction of applied strain, and
$S(\varepsilon)=\diag(c_\parallel , c_\perp )$ the matrix describing the deformation of the
Dirac cone. Explicitly, for the compound transformation matrix $R(\theta)
S(\varepsilon) R(-\theta)$ mapping $\bq$ onto $\bq^\prime$ one finds 
\begin{equation}
R(\theta) S(\varepsilon) R(-\theta) = {\mathbb I} -
2\kappa{\boldsymbol\varepsilon}.
\end{equation}

A central result of the present work is that a similar correspondence holds
between a generic linear response function $\chi (\bq,\omega)$ under applied
strain, with respect to its unstrained limit, $\chi^{(0)} (\bq,\omega)$. This
follows from the fact that any linear response function $\chi (\bq,\omega)$ of a
noninteracting electron system can be expressed as an integral over the first
Brillouin zone (1BZ) of a suitable matrix operator over pseudospins, which is
itself a function of $\bq$. Such an operator then admits a unique expression in
terms of the Pauli matrices $\sigma_x$, $\sigma_y$, $\sigma_z$, and the identity
matrix ${\mathbb I} \equiv \sigma_0$. The simplest cases are then given by the
density operator and the density current operator, which in reciprocal space
read \cite{Principi:09}
\begin{subequations}
\begin{eqnarray}
\rho^{(0)} (\bq)&=& \sum_\bk \Psi_{\bk-\bq}^\dag {\mathbb I} \Psi_\bk , \\
J^{(0)}_i (\bq) &=& -e\vF \sum_\bk \Psi_{\bk-\bq}^\dag \sigma_i \Psi_\bk , \quad
i=x,y,
\end{eqnarray}
\end{subequations}
respectively, where $\Psi_\bq^\dag = (\psi_{\bq A} , \psi_{\bq B}
)$, and $\psi_{\bq\alpha}$ destroys a quasiparticle with momentum $\bq$ and
pseudospin $\alpha=A,B$, and summations run over the 1BZ. While the
density operator does not change under applied strain, for the generic component
of the density current operator one has 
\begin{equation}
J_i = [{\mathbb I} - 2\kappa{\boldsymbol\varepsilon}]_{ij}
J_j^{(0)}.
\end{equation}
Here and below a summation will be understood over repeated indices ($j=x,y$). 

Defining now eigenvalues and eigenvectors in pseudospin space of the Hamiltonian
with and without applied strain, Eqs.~(\ref{eq:Hunstrained}) and
(\ref{eq:Hstrained}), as $H^{(0)} |\bq^\prime , \lambda\rangle^{(0)} =
E_{\lambda\bq^\prime}^{(0)} |\bq^\prime , \lambda\rangle^{(0)}$ and $H |\bq ,
\lambda\rangle = E_{\lambda\bq} |\bq , \lambda\rangle$, respectively, with
$\lambda$ a pseudospin index, it follows that both $E_{\lambda\bq}$ and $|\bq ,
\lambda\rangle$ under applied strain are mapped onto
$E_{\lambda\bq^\prime}^{(0)}$ and $|\bq^\prime , \lambda\rangle^{(0)}$,
respectively, where $\bq^\prime$ is given in terms of $\bq$ by
Eq.~(\ref{eq:straintransform}). Performing such a linear change of variables in
the integral defining the correlation function of interest, in the cases of the
density-density and current-current correlation function, it follows therefore
that
\begin{subequations}
\label{eq:chi}
\begin{eqnarray}
\label{eq:chi:scalar}
\Pi_{\rho\rho} (\bq,\omega) &=& \left[\det S(\varepsilon)\right]^{-1} \Pi_{\rho\rho}^{(0)}
(\bq^\prime ,\omega) , \\
\Pi_{ij} (\bq,\omega) &=& \left[\det S(\varepsilon)\right]^{-1} \nonumber\\
\label{eq:chi:vector}
&&\times
[{\mathbb I} - 2\kappa{\boldsymbol\varepsilon}]_{ih}
\Pi_{hk}^{(0)} (\bq^\prime , \omega)
[{\mathbb I} - 2\kappa{\boldsymbol\varepsilon}]_{kj} ,
\end{eqnarray}
\end{subequations}
where $\det S(\varepsilon) = (1-2\kappa\varepsilon)(1+2\kappa\nu\varepsilon)$.
From Eq.~(\ref{eq:chi:scalar}), in the case of the density-density correlation
function, it follows in particular that the effect of applied strain is that of
transforming the momentum variable $\bq$ into an `effective' one $\bq^\prime$,
plus the introduction of an overall scale factor $\left[\det
S(\varepsilon)\right]^{-1}$, which is isotropic with respect with the strain
direction. Such a scale factor is directly related to the slope of the
electronic density of states at the Fermi level. As is well known, this goes
linearly with the chemical potential $\mu$, and it has been shown that its
steepness increases with increasing strain, for moderately low strain
modulus \cite{Pellegrino:09b}. In the case of the current-current correlation
function, such an overall effect is then superimposed to an anisotropic
deformation, depending on the angle of applied strain, $\theta$, as shown by
Eq.~(\ref{eq:chi:vector}).

Linearizing Eq.~(\ref{eq:chi}) with respect to $\varepsilon$, one finds
\begin{subequations}
\label{eq:chilinear}
\begin{eqnarray}
\label{eq:chilinear:scalar}
\Pi_{\rho\rho} (\bq,\omega) &=& \left[ 1 + 2\kappa(1-\nu)\varepsilon\right] \Pi_{\rho\rho}^{(0)}
(\bq,\omega) \nonumber\\
&&- 2\kappa \frac{\partial \Pi_{\rho\rho}^{(0)} (\bq,\omega)}{\partial q_h}
\varepsilon_{hk} q_k , \\
\label{eq:chilinear:vector}
\Pi_{ij} (\bq,\omega) &=& \left[ 1 + 2\kappa(1-\nu)\varepsilon\right] \Pi_{ij}^{(0)}
(\bq,\omega) \nonumber\\
&&- 2\kappa \frac{\partial \Pi_{ij}^{(0)} (\bq,\omega)}{\partial q_h}
\varepsilon_{hk} q_k \nonumber\\
&&-2\kappa \{ {\boldsymbol\varepsilon}, {\boldsymbol\Pi}^{(0)} (\bq,\omega) \}_{ij} ,
\end{eqnarray}
\end{subequations}
where the curly brackets in the last term denote a matrix anticommutator.

\section{Polarization}
\label{sec:polarization}

\subsection{Charge response: Plasmons and conductivity}

We now specifically turn to consider the density-density correlation function
within linear response theory, \emph{i.e.} the electron polarization
$\Pi_{\rho\rho} (\bq,\omega)$. Plasmon modes are then recovered as poles of the
polarization, and the effect of strain on their dispersion relation has been
studied in Refs.~\onlinecite{Pellegrino:10a,Pellegrino:10c}. In particular, by
including local field effects, as a consequence of the two-band character of the
band structure of graphene, we found two main plasmon branches, the lower branch
being characterized by the standard square-root dependence on $q$, at long
wavelengths, as expected for a two-dimensional system \cite{Pellegrino:10a}.

In given limits, the asymptotic form of the noninteracting polarization in the
absence of strain, say $\Pi_{\rho\rho}^{(0)} (\bq,\omega)$ is known explicitly. For
instance, in the long wavelength limit ($q\to0$), one finds \cite{Wunsch:06}
\begin{eqnarray}
\Pi_{\rho\rho}^{(0)} (q\to0,\omega) &=& \frac{g_s g_v q^2}{8\pi\hbar\omega} \left[
\frac{2\mu}{\hbar\omega} + \frac{1}{2} \log \left|
\frac{2\mu-\hbar\omega}{2\mu+\hbar\omega} \right| \right. \nonumber\\
&&\left. -i\frac{\pi}{2}
\Theta(\hbar\omega-2\mu) \right] ,
\label{eq:Wunsch}
\end{eqnarray}
where $g_s =g_v =2$ take into account for spin and valley degeneracies,
respectively. In other words, $\Pi_{\rho\rho}^{(0)} (q\to0,\omega) =
Z(\omega) q^2$, at a given $\omega$, with the complex factor $Z(\omega)$
implicitly defined by Eq.~(\ref{eq:Wunsch}).

In the case of applied strain, but still in the noninteracting limit,  this is
then readily modified through the linearized Eq.~(\ref{eq:chilinear}), yielding
\begin{equation}
\Pi_{\rho\rho} (q\to0,\omega) = \left[ 1 - 2\kappa(1+\nu) \varepsilon \cos
(2\theta-2\varphi) \right] Z(\omega) q^2 ,
\end{equation}
where $\bq \equiv q (\cos\varphi,\sin\varphi)$. Within the random phase
approximation (RPA), the interacting polarization reads  $\bar{\Pi}_{\rho\rho}
(\bq,\omega)= \Pi_{\rho\rho} (\bq,\omega)/ (1-V(q) \Pi_{\rho\rho}
(\bq,\omega))$, where $V(q) = e^2 / (2\epsilon_r \epsilon_0 q)$ is the (bare)
Coulombic electron-electron interaction, and $\epsilon_r$ is the dielectric
constant of the medium. Solving for the plasmon dispersion relation, $\Re
\bar{\Pi}_{\rho\rho}^{-1} (\bq,\omega) = 0$, at low energies one finds
\begin{eqnarray}
\hbar\omega_{\mathrm{pl}} &=& \sqrt{\frac{e^2}{2\pi\epsilon}\mu}
\left[ 1 - \kappa(1+\nu)\varepsilon \cos(2\theta-2\varphi)\right] \sqrt{q}
\nonumber\\
&\equiv& \hbar\tilde{\omega}_1 (\varphi) \sqrt{qa} .
\label{eq:omegapl}
\end{eqnarray}
One thus finds that the prefactor $\tilde{\omega}_1 (\varphi)$ in the
$\sqrt{q}$-dependence is maximum [resp., minimum] for $\varphi-\theta=\pi/2$
[$\varphi-\theta=0$], \emph{i.e.} wavevector orthogonal [parallel] to the
direction of applied strain. Correspondingly, one also finds for the imaginary
part of the retarded polarizability along the low-energy plasmon branch 
\begin{eqnarray}
\Im \bar{\Pi}_{\rho\rho} (\bq,\omega+i0^+ ) &=& \nonumber\\
&&\hspace{-2truecm}
-\frac{1}{2} \sqrt{\frac{2\pi\epsilon}{e^2} \mu} 
\left[ 1 - \kappa(1+\nu)\varepsilon \cos(2\theta-2\varphi)\right] \nonumber\\
&&\hspace{-2truecm}
\times (qa)^{3/2}
\delta(\hbar\omega - \hbar\omega_{\mathrm{pl}}(\bq)) .
\end{eqnarray}
Therefore, one recovers a dependence of the plasmon spectral weight on the angle
of applied strain, similar to that shown by $\tilde{\omega}_1 (\varphi)$ in
Eq.~(\ref{eq:omegapl}).

Another quantity of interest which is related to the density-density correlation
function is the optical conductivity, which can be obtained as
\begin{equation}
\sigma_{\varphi\varphi} (\omega) = \frac{ie^2}{\omega} \lim_{q\to0}
\frac{\omega^2}{q^2} \Pi_{\rho\rho} (\bq,\omega) .
\end{equation}
Making use of Eq.~(\ref{eq:chilinear}) and (\ref{eq:Wunsch}) one therefore finds
the optical conductivity in the presence of applied strain as
\begin{eqnarray}
\sigma_{\varphi\varphi} (\omega) &=& \sigma_0 \left[ 1
-2\kappa(1+\nu)\varepsilon\cos(2\theta-2\varphi)\right] \nonumber\\
&&\hspace{-1.5truecm}\times
\left(
\Theta(\hbar\omega-2\mu)+ i \frac{4}{\pi}\frac{\mu}{\hbar\omega} + \frac{i}{\pi}
\log \left| \frac{2\mu-\hbar\omega}{2\mu+\hbar\omega} \right| \right) ,
\label{eq:sigma}
\end{eqnarray}
where $\sigma_0 = \pi e^2 / 2h$ is proportional to the quantum of conductivity.
In the hydrostatic limit, $\nu=-1$, $\sigma_{\varphi\varphi}$ does not depend on
strain, as may be expected, as the unstrained relation does not contain the
Fermi velocity.

The above expression for the conductivity, Eq.~(\ref{eq:sigma}), can be
exploited to study the strain dependence of the transverse electromagnetic mode,
that has been recently predicted theoretically in graphene \cite{Mikhailov:07},
and in a graphene bilayer \cite{Jablan:11a}.
In a 2D electron gas, the spectrum of electromagnetic modes obeys the equations
\cite{Falko:89}
\begin{subequations}
\begin{eqnarray}
\label{eq:omega-long}
1+ i\frac{\sigma}{2\epsilon_0 \omega} \zeta (q,\omega) &=& 0, \\
\label{eq:omega-trans}
1-i \frac{\sigma}{2\epsilon_0} \frac{\omega}{\zeta (q,\omega) c^2} &=& 0 ,
\end{eqnarray}
\end{subequations}
for the longitudinal and transverse plasmons, respectively, where $\zeta^2
(q,\omega) = q^2 - (\omega/c)^2$, where $c$ is the velocity of light in vacuum.
While conventional 2D electron systems cannot sustain a transverse
electromagnetic mode, it has been predicted \cite{Mikhailov:07} that graphene
can develop a transverse plasmon mode, as a consequence of a negative imaginary
part in the interband contribution to its optical conductivity,
Eq.~(\ref{eq:sigma}). Its logarithmic divergence as $\hbar\omega/\mu\to 2$ is in
turn related to the discontinuous behavior of the interband absorption of
radiation at frequencies $\hbar\omega > 2\mu$.
Such a feature is a generic consequence of causality, and is related through a
Kramers-Kr\"onig transformation to the step-like behaviour of the real part of
the optical conductivity. This is in turn due to the existence of a Fermi
surface, which is however expected to be smeared at finite temperature, thus
implying the reduction of the logarithmic singularity into a pronounced (but
finite) peak.

Indeed, making use of Eq.~(\ref{eq:sigma}) in Eq.~(\ref{eq:omega-long}), one
consistently recovers Eq.~(\ref{eq:omegapl}) for the longitudinal plasmons. On
the other hand, substituting Eq.~(\ref{eq:sigma}) in Eq.~(\ref{eq:omega-trans}),
one obtains the strain-dependence of the dispersion relation of the transverse
plasmon implicitly as
\begin{eqnarray}
\frac{\hbar c}{\alpha\mu} \zeta (q,\omega) &=& \left( 1 -2 \kappa (1+\nu)\varepsilon
\cos(2\theta-2\varphi)\right) \nonumber\\
&&\times \left[ \frac{\hbar\omega}{2\mu} \log \left|
\frac{2\mu+\hbar\omega}{2\mu-\hbar\omega} \right| -2 \right],
\label{eq:omega-trans-strain}
\end{eqnarray}
where $\alpha=e^2/(4\pi\epsilon_0 \hbar c)$ is the fine structure constant.

\begin{figure}[t]
\centering
\includegraphics[height=0.8\columnwidth,angle=-90]{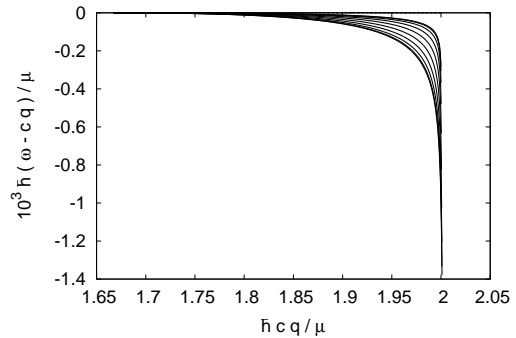}
\caption{Showing deviations from linearity of the frequency of the transverse
plasmon, Eq.~(\ref{eq:omega-trans-strain}), for $q$ in the allowed range, for
strain modulus $\varepsilon=0.1$, and strain direction ranging from
$\varphi-\theta=0$ (top) to $\varphi-\theta=\pi/2$ (bottom).}
\label{fig:omega-trans-strain}
\end{figure}

Because of the small factor $\alpha$ in the left-hand side of
Eq.~(\ref{eq:omega-trans-strain}), the dispersion relation of such a transverse
mode is close to the linear dispersion relation of the
electromagnetic radiation itself, $\omega -cq \lesssim 0$. However, one may
expect that applied strain enhances deviations from linearity
(\emph{i.e.,} from the photon's dispersion relation), as a
consequence of a strain-induced modification of the band dispersion.
Fig.~\ref{fig:omega-trans-strain} shows indeed deviations from linearity,
$\omega -cq$ of the transverse plasmon, for $q$ in the allowed range, for strain
modulus $\varepsilon=0.1$, and strain direction $0\leq \varphi-\theta\leq\pi/2$.
One finds indeed that, in the case of applied strain, deviations of the
transverse plasmon dispersion relation from that of the electromagnetic
radiation become significant over a sufficiently wide window in $\hbar c q/\mu
\lesssim 2$, especially when $\varphi-\theta=\pi/2$.

Therefore, applied strain should help the experimental detection of
this elusive collective mode. Indeed, the fact that the plasmon dispersion
relation is close to the corresponding electromagnetic dispersion implies that
such a transverse plasmon mode would have a marked photonic character, and a
small plasmon linewidth would therefore hinder its observation \cite{Jablan:11}.
On the other hand, at finite temperature, the real part of the optical
conductivity is nonzero also for $\hbar\omega\lesssim 2\mu$, so that the
transverse plasmon mode does acquire a finite, albeit small, linewidth
\cite{Mikhailov:07}. In particular, this applies to plasmon energies such that
$0<2\mu-\hbar\omega<\kB T$. This is exactly where the plasmon dispersion
relation deviates most from its photonic counterpart, the deviation being
enhanced, and shifted away from the limiting case $\hbar\omega=2\mu$, in the
case of applied strain, for $\bq$ perpendicular to the strain direction. One
therefore expects wavevectors of the order of $\hbar cq \lesssim 2\mu$, or
equivalently $q/\kF \lesssim 2\vF/c \ll 1$, so that it is justified to employ
Eq.~(\ref{eq:omega-trans-strain}) \cite{Wunsch:06,Mikhailov:07}.

\subsection{Current response}

In the case of an applied vector field (\emph{e.g.,} an electric field $E_i$),
one may in general decompose the linear response function in a longitudinal and
a transverse component as 
\begin{equation}
\chi_{ij}(\bq,\omega) = \frac{q_i q_j}{q^2} \chi_\parallel (q,\omega) + \left(
\delta_{ij} - \frac{q_i q_j}{q^2} \right) \chi_\perp (q,\omega) ,
\end{equation}
where $q=|\bq|$, for a homogeneous system \cite{Giuliani:05}. In particular, in
the case of the current-current correlation function, the latter being
proportional to the pseudospin-pseudospin counterpart, this can be further
simplified as
\begin{equation}
\Pi_{ij}^{(0)} (\bq,\omega) = \Pi^{(0)}_+ (q,\omega) \delta_{ij}
+ \Pi^{(0)}_- (q,\omega) A_{ij} (\varphi),
\end{equation}
where
\begin{equation}
\Pi^{(0)}_\pm (q,\omega) = \frac{1}{2}[\Pi^{(0)}_\parallel (q,\omega) \pm
\Pi^{(0)}_\perp (q,\omega)].
\end{equation}
Therefore, as a consequence of Eq.~(\ref{eq:chi}), it follows that even though
an unstrained system is characterized only by transverse response in
the static limit \cite{note:nolongitudinal}, it may develop a nonzero parallel
response as a result of a strain-induced deformation. Making use of
Eq.~(\ref{eq:chilinear:vector}), one finds
\begin{widetext}
\begin{eqnarray}
\Pi_{ij}(\bq,\omega) &=& \Pi_{ij}^{(0)} (\bq,\omega)
-2\varepsilon\kappa(1+\nu) \left[
\Pi_-^{(0)} (q,\omega)
\cos(2\theta-2\varphi)\delta_{ij} + \Pi_+^{(0)} (q,\omega)
A_{ij}(\theta) + \Pi_-^{(0)} (q,\omega) A_{ij} (\varphi+\pi/4)
\sin(2\theta-2\varphi)\right] \nonumber\\
&&-\kappa [(1-\nu)+(1+\nu)\cos(2\theta-2\varphi)]\varepsilon\left[
q\frac{\partial\Pi_+^{(0)}(q,\omega)}{\partial q} \delta_{ij}
+
q\frac{\partial\Pi_-^{(0)}(q,\omega)}{\partial q} A_{ij} (\varphi)
\right].
\label{eq:Piij}
\end{eqnarray}
In the static limit ($\omega=0$), Eq.~(\ref{eq:Piij})
can be further simplified, by considering the analytic result of
Ref.~\onlinecite{Principi:09}, with $\Pi^{(0)}_\parallel (\bq,0)=0$, and
\begin{equation}
\Pi^{(0)}_\perp (\bq,0) = \frac{g_s g_v e^2 \vF}{16\hbar q} 
\left[1-\Theta \left(\hbar \vF q - 2 \mu \right)\frac{2}{\pi} \left[\arcsin \left(\frac{2 \mu}{\hbar \vF q}\right)-
\frac{2 \mu}{\hbar \vF q}\sqrt{1-\left(\frac{2 \mu}{\hbar \vF q}\right)^2}\right]
-\Theta \left(2 \mu -\hbar \vF q \right)
\right].
\label{eqn:transv}
\end{equation}
\end{widetext}
In particular, one
recovers a vanishing response, $\Pi_{\bq\bq} (\bq,0)=0$, with
$\Pi_{\bq\bq}$ denoting the current-current correlation function for both vector
potential and response field aligned with $\bq$, when $\bq$ is aligned with the
applied field also in the presence of strain, as expected in the static limit.

\section{Electric and magnetic susceptibilities}
\label{sec:susceptibilities}

A magnetic field applied in the direction perpendicular to the graphene plane
can be described as ${\mathbf B}_{\mathrm ext} (\bq) = B_{\mathrm ext} (\bq)
\hat{\mathbf z} = i \bq\times{\mathbf A}$, where ${\mathbf A} = i (q_y , -q_x )
B_{\mathrm ext} /q^2$, in reciprocal space. The linear response to such a
magnetic field is then given by a current $J_i$, which in turn produces a
magnetization term $\delta{\mathbf B} \equiv\chi_m {\mathbf B}_{\mathrm ext}$.

In the case of a static, uniform magnetic field, oriented in the direction
orthogonal to the graphene sheet, one is interested in the magnetic
susceptibility defined as
\begin{equation}
\chi_M=\lim_{q \to 0} \int \frac{d \varphi}{2 \pi} \chi_m (\bq,0).
\end{equation}
Making use of Eq.~(\ref{eqn:transv}), one obtains
\begin{equation}
\chi_M=\lim_{q \to 0} \left( -\frac{\mu_0}{q^2} \right) \left[1 -\kappa(1-\nu)\varepsilon q
\frac{\partial }{\partial q}\right]\Pi^{\perp(0)} (q,0).
\label{eq:chim}
\end{equation}
In the strained case, this reads
\begin{equation}
\chi_M=-\mu_0[1-2 \kappa(1-\nu)\varepsilon ] \frac{g_s g_v e^2 v_F^2}{6 \pi}
\delta(\mu).
\label{eq:chim0}
\end{equation}
One therefore obtains a qualitatively similar result to the case of undeformed
graphene, treated within the Dirac approximation and neglecting the
electron-electron interaction \cite{McLure:56,Principi:09}. On the other hand,
applied strain causes a reduction of the magnetic response, Eq.~(\ref{eq:chim}).
Although Eq.~(\ref{eq:chim0}) would imply no response to a static, uniform
magnetic field away from half-filling, one expects that finite-temperature
effects would broaden the $\delta$-function, already in the noninteracting
limit. A qualitatively similar smearing of the peak in the dependence on the
chemical potential may also be induced by disorder \cite{Koshino:07}. Still at
zero temperature and in the noninteracting limit, one recovers a nonzero
magnetic response also away from half-filling, when the honeycomb lattice
structure is considered \cite{GomezSantos:11}. The effect of the interactions
has been considered in Ref.~\onlinecite{Principi:10}, where it is shown that an
interacting 2D Dirac electron liquid develops a magnetic response also at finite
doping.

An analogous procedure may be followed to derive the electric susceptibility
$\chi_e$, entering the relationship $\delta{\mathbf E} = \chi_e {\mathbf
E}_{\mathrm ext}$ between the electric polarization and an external electric field. One
is then interested in the static ($\omega=0$) limit of the density-density
polarization. In the presence of applied strain, at arbitrary $\mu=\hbar\vF\kF$, using
Eq.~(\ref{eq:chilinear}), one explicitly finds
\begin{widetext}
\begin{eqnarray}
\Pi_{\rho\rho}(q,\omega=0) &=& \left[ 1 + 2  \kappa (1-\nu) \varepsilon\right]
\left[
-\frac{g_s g_v \mu}{2\pi\hbar^2 \vF^2} + \frac{g_s g_v q}{8\pi\hbar \vF}
G^+_< \left( \frac{2\mu}{\hbar \vF q} \right) \Theta(\hbar \vF q-2 \mu)
\right] \nonumber\\
&&-\kappa [(1-\nu)+(1+\nu)\cos(2\theta-2\varphi)] \varepsilon
\frac{g_s g_v q}{8\pi\hbar \vF}
G^-_< \left( \frac{2 \mu}{\hbar \vF q} \right) \Theta(\hbar \vF q-2 \mu),
\label{eq:prechie}
\end{eqnarray}
\end{widetext}
where \cite{Wunsch:06,Stauber:10a}
\begin{equation}
G^\pm_< (x) = \pm x \sqrt{1-x^2} - \arccos x, \quad |x|<1 .
\end{equation}
In particular, at zero doping ($\mu=0$, $G^\pm_< (0) = -\pi/2$),
one finds in general that
\begin{eqnarray}
\chi_e (\bq,0) &=& V(q) \Pi_{\rho\rho} (q,0)  .
\label{eq:chie}
\end{eqnarray}

It should be emphasized that, while Eq.~(\ref{eq:chie}) describes the response
of the system to a static electric field lying in the same graphene layer.
More explicitly, in the undoped case, Eq.~(\ref{eq:chie}) reads
\begin{eqnarray}
\chi_e= \lim_{q \to 0} \chi_e ({\mathbf q},0) &=& 
-\frac{g_s g_v e^2}{32\epsilon_0 \epsilon_r \hbar \vF} \nonumber\\
&&\hspace{-2.5truecm}
\times \left[
1 + \kappa(1-\nu)\varepsilon -
\kappa(1+\nu)\varepsilon\cos(2\theta-2\varphi) \right] ,
\label{eq:chieexpl}
\end{eqnarray}
where $\varphi$ is the direction of the electric field on the graphene plane,
thus showing that uniaxial strain introduces a modulation in the angle of
applied strain. Moreover, comparison with Eq.~(\ref{eq:chim}) in the hydrostatic
limit ($\nu=-1$) shows that strain enhances the electric response, while
suppressing the magnetic one.

\section{Conclusions}
\label{sec:conclusions}

We have studied the dependence on applied uniaxial strain of several
linear-response electronic correlation functions of graphene. After deriving a
general correspondence between strained and unstrained correlation functions, we
have derived the strain dependence of the plasmon dispersion relation and of the
optical conductivity. Specifically, we find that the prefactor in the
$\sqrt{q}$-dependence of the plasmon frequency develops an anisotropic
character, with maxima occurring when the wavevector is orthogonal to the
direction of applied strain. Moreover, we derive a strain-induced anisotropic
enhancement of the deviations from linearity of the recently predicted
transverse plasmon \cite{Mikhailov:07}, which should facilitate its experimental
detection in suitably strained graphene samples. Finally, we have compared and
contrasted the strain dependences of the magnetic and electric susceptibilities,
showing that strain enhances the response of strained graphene to an applied
electric field, while suppressing the response to an magnetic field.

\acknowledgments

FMDP acknowledges Dr D. M. Basko for inspiring discussions, and for carefully
reading the manuscript. FMDP also thanks the Laboratoire de Physique et
Mod\'elisation des Mileux Condens\'es, CNRS, Grenoble (France) for much
hospitality and for the stimulating environment.

\begin{small} \bibliographystyle{apsrev}
\bibliography{a,b,c,d,e,f,g,h,i,j,k,l,m,n,o,p,q,r,s,t,u,v,w,x,y,z,zzproceedings,Angilella,notes}

\begin{thebibliography}{47}
\bibitem[{not({\natexlab{b}})}]{note:nolongitudinal}
\bibinfo{note}{The static ($\omega\to0$) longitudinal response describes the
  response of the system to a static longitudinal vector potential, which can
  always be removed via a gauge transformation. Therefore, such a contribution
  to the linear response must be zero. See also Ref.~\onlinecite{Giuliani:05}.}
\end{thebibliography}
\end{small}

\end{document}